\documentclass[12pt,a4paper]{article}

\usepackage{epsfig}
\usepackage{amsmath,amsfonts,amssymb}
\usepackage{t1enc}
\usepackage{cite}
\usepackage{colortbl}
\usepackage{pifont}
\usepackage{slashed}
\usepackage{multirow}
\usepackage{bbold}
\usepackage[list=true]{subcaption}

\providecommand{\openone}{\leavevmode\hbox{\small1\kern-3.8pt\normalsize1}}

\def\Tr{\mbox{Tr}}

\usepackage{colortbl,colordvi,xcolor,color}
\usepackage{rotating}

\usepackage[T1]{fontenc} 

\newcommand{\ilog}{I_{log}(\lambda^{2})}

\newcommand{\Log}[1]{\log\left(-\frac{{#1}^{2}}{\lambda^{2}}\right)}
\newcommand{\bb}[1]{\bar{#1}}

\newcommand{\beq}{\begin{equation}}
\newcommand{\eeq}{\end{equation}}
\newcommand{\bea}{\begin{eqnarray}}
\newcommand{\eea}{\end{eqnarray} }
\newcommand{\nn}{\nonumber \\}

\newcommand{\ga}{\gamma}

\newcommand{\lag}{\mathcal L}

\begin{document}

\begin{center}
\begin{Large}
{\bf Step towards a consistent treatment of chiral theories at higher loop order: the abelian case}
\end{Large}

\vspace{0.5cm}
Adriano\ Cherchiglia  \\[0.2cm] 
{ \it 
Instituto de Física Gleb Wataghin - UNICAMP, 13083-859, Campinas SP, Brazil}
\end{center}

\begin{abstract}
As recently pointed out, regularization schemes defined in four-dimensions may also face inconsistencies in the presence of chiral fermions. In this work, we extend this analysis to two-loop order. Adopting the implicit regularization as working arena, 
we discuss in detail how a consistent version of the method can be envisaged, and present many of the subtleties  that appear at two-loop order using an abelian chiral model as working example.
\end{abstract}

\section{Introduction}
\label{sec:intro} 

In view of the amount of data already collected at LHC as well as the High-Luminosity program to be carried in the following years~\cite{Strategy:2019vxc}, it is of utmost importance that theoretical predictions are known at NNLO and beyond (see \cite{Heinrich:2020ybq} for a recent review). In order to tackle this challenge, innovative techniques have been proposed in recent years, aiming to perform the subtraction of divergences at integrand level as well as staying as long as possible in the physical dimension. Techniques of this kind are generally denoted regularization methods in fixed dimension as opposed to more traditional methods, for instance, dimensional regularization \cite{Bollini:1972ui,tHooft:1972tcz}.

From a practical point of view, regularization methods in fixed dimensions may be more prone to numerical approaches since the UV divergent parts of generic Feynman amplitudes can be disentangled in a purely algebraic way. Since dimensional methods, in general, require the analytical treatment of integrals to identify the coefficients of the terms in a series expansion in $\epsilon$, these alternative methods may be more efficient by minimizing the analytical treatment. This is one of the motivations to study and further develop alternative regularization techniques.
 For a review, we refer the reader to the references \cite{Gnendiger:2017pys,TorresBobadilla:2020ekr}. 

On the other hand, one of the main hopes of methods defined in fixed dimension was to deal with dimension-specific objects (the $\gamma_{5}$ matrix as the most prominent example) in a straightforward way, with no ambiguities. However, as showed in a previous work \cite{Bruque:2018bmy}, this expectation was too naive, meaning that these methods share similar problems with $\gamma_{5}$ as dimensional ones (see \cite{Jegerlehner:2000dz} for a review). The way out was to define them in a quasi-dimensional space, such that the genuine space is embed in it. This is similar to the construction used in the four-dimensional helicity scheme (FDH) and dimensional reduction (DRED) methods \cite{Siegel:1979wq,Siegel:1980qs,Kilgore:2011ta,Jack:1994bn,Jack:1993ws,Signer:2005iu}. However, the approach in the context of methods in fixed dimension is simpler since only two spaces must be explicitly defined as opposed to four in the case of FDH/DRED. Nevertheless, by proposing to define methods in fixed dimension in a quasi-dimensional space we are actually adopting a scheme similar to the Breitenlohner-Maison-'t Hooft-Veltman (BMHV) scheme, which was originally conceived for dimensional regularization \cite{Breitenlohner:1975hg,Breitenlohner:1977hr,Breitenlohner:1976te}. This claim was already made in \cite{Bruque:2018bmy}, however the examples provided there were restricted to one-loop order. The main aim of this contribution is to extend the analysis to NNLO, providing a step towards a consistent method in fixed dimension to tackle chiral theories without ambiguities to arbitrary loop order.  As in the case of the BHMV scheme, one drawback to be faced is the breaking of gauge invariance in intermediate steps, which requires the inclusion of finite symmetry-restoring counterterms. An extensive study of the counterterms required at one-loop in the context of dimensional regularization was recently performed \cite{Belusca-Maito:2020ala}, which complements previous ones in the literature, for instance \cite{Martin:1999cc,SanchezRuiz:2002xc}. For the specific examples we consider, we will reproduce the results of those references in the context of methods in fixed dimension as well. Recently, the analysis of \cite{Belusca-Maito:2020ala} was also extended to two-loop order \cite{Belusca-Maito:2021lnk}.


Our work is organized as follows: in section \ref{sec:ireg} we present an example of method in fixed dimension denoted Implicit Regularization, presenting the main results and ideas needed in subsequent sections. We discuss how the method should be implemented in a way to be consistent, even in the presence of chiral fermions. In section \ref{sec:QED} we provide an abelian chiral theory to discuss in detail the issues that must be faced when moving to two-loop order, such as the inclusion of symmetry-restoring counterterms as well as the renormalization of the theory. Finally, we conclude in section \ref{sec:conclusions}.

%
%
%

\section{The IREG formalism}
\label{sec:ireg} 

The implicit regularization method, first presented in \cite{Battistel:1998sz}, is based on a very simple premise: the UV divergent part of a general Feynman diagram can always be expressed in terms of integrals void of physical parameters (masses/external momenta)\footnote{The same idea is the basis of the Four-dimensional regularisation/renormalisation (FDR) method, later proposed in \cite{Pittau:2012zd}.}. In order to implement this premise, a mathematical identity is applied at integrand level allowing a separation between a UV only divergent integral (independent of external momenta/masses) and a finite part. An important point to be noticed is that spurious IR divergences are generated in intermediate steps, which requires the inclusion of a  $\mu$ mass IR regulator. Being IR safe from the start\footnote{When IR divergences do appear, the same $\mu$ will play the role of an IR regulator, canceling in IR-safe observables \cite {Gnendiger:2017pys,TorresBobadilla:2020ekr}.}, the limit $\mu\rightarrow 0$ is well-defined for the Feynman amplitude as a whole. Moreover, it provides a natural way to add a renormalization scale, which we denote by $\lambda$. From the point of view of applications, IREG were employed, for instance, in processes such as $h\rightarrow \gamma\gamma$ \cite{Cherchiglia:2012zp}, and $e^{-}e^{+}\rightarrow \gamma^{*} \rightarrow q\bar{q}(g)$ \cite{Gnendiger:2017pys}. Moreover, it was used to study the coupling beta function in scalar \cite{Cherchiglia:2010yd}, abelian and non-abelian \cite{Cherchiglia:2020iug,Cherchiglia:2021yxz}, as well as supersymmetric \cite{Cherchiglia:2015vaa,Fargnoli:2010mf,Carneiro:2003id} theories. Finally, it was shown that the method respects abelian gauge-symmetry to arbitrary loop order \cite{Vieira:2015fra,Ferreira:2011cv}. For further applications, see \cite{Arias-Perdomo:2021inz} and references therein.

After this brief overview, we provide an 1-loop example to illustrate the procedure. Consider the scalar self-energy in the massless $\phi^3$ theory whose amplitude is given by
\begin{align}
\Xi^{(1)}\equiv\frac{g^{2}}{2}\int\limits_{k}\!\frac{1}{k^{2}}\frac{1}{(k-p)^{2}}=\lim_{\mu^{2}\rightarrow 0}\frac{g^{2}}{2}\int\limits_{k}\!\frac{1}{(k^{2}-\mu^{2})}\frac{1}{[(k-p)^{2}-\mu^2]},\quad \int\limits_{k}\equiv\int\frac{d^4 k}{(2 \pi)^4}.
\end{align}
where $p$ is the external momenta and in the second equation we have added an IR mass regulator $\mu^2$ to control spurious intermediate IR divergences. 
In order to obtain an UV divergent part independent of the external momentum we apply the identity below as many times as necessary
\begin{align}
	\frac{1}{(k-p)^2-\mu^2}=\frac{1}{(k^2-\mu^2)}
	-\frac{(p^2-2 \cdot k)}{(k^2-\mu^2)
		\left[(k-p)^2-\mu^2\right]},
	\label{ident}
\end{align}  
amounting to
\begin{align}
\Xi^{(1)}=\lim_{\mu^{2}\rightarrow 0}\frac{g^{2}}{2}\left[\int\limits_{k}\!\frac{1}{(k^{2}-\mu^{2})^{2}}-\int\limits_{k}\!\frac{(p^2-2 \cdot k)}{(k^2-\mu^2)^{2}\left[(k-p)^2-\mu^2\right]}\right].
\end{align}

At this point, one can notice that the UV divergent part will also develop an IR divergence after taking the limit $\mu^2\rightarrow 0$. This behavior is undesirable, being avoided by the introduction of another parameter ($\lambda>0$) which will play the role of the renormalization group scale. To this end, we rewrite the UV divergent term as
\beq
\int\limits_{k}\frac{1}{(k^{2}-\mu^{2})^{2}}=\int\limits_{k}\frac{1}{(k^{2}-\lambda^{2})^{2}}+ b\ln \frac{\lambda^2}{\mu^2},\quad b\equiv\frac{i}{(4 \pi)^2},
\label{scale}
\eeq 
which disentangles the UV divergent dependence (encoded by the integral dependent on $\lambda$) from the spurious IR divergence (parametrized as the logarithm). The limit $\mu^2\rightarrow 0$ can now be safely performed and we finally obtain
\begin{align}
\Xi^{(1)}=\frac{g^{2}}{2}\left[\int\limits_{k}\!\frac{1}{(k^{2}-\lambda^{2})^{2}}-b\ln\frac{p^2}{\lambda^2}+2b\right].
\end{align}

Although this example is very simple, it shows a generic behavior of IREG: the UV divergent part of a general 1-loop Feynman amplitude will always be proportional to
\bea
	I_{log}(\lambda^2)&\equiv& \int_{k} \frac{1}{(k^2-\lambda^2)^{2}}.
	\eea

By performing subtractions of $\ilog$ one defines a minimal subtraction scheme in the context of IREG, which is mass-independent by construction. Thus, 1-loop renormalization functions $Z_{i}$ will be proportional to $\ilog$ in general. Since derivatives of $Z_{i}$ are necessary ingredients when obtaining renormalization group functions, derivatives of $\ilog$ will be needed
\beq
\lambda^2\frac{\partial I_{log}(\lambda^2)}{\partial \lambda^2}= -b.
\label{eq:derivative}
\eeq

At higher loop order a similar program can be envisaged in a way compatible with Bogoliubov's recursion formula \cite{Cherchiglia:2010yd}. Since we are adopting a mass-independent subtraction scheme, the UV divergent part will only contain a logarithmic dependence, allowing us to define
\begin{align}
I_{log}^{(l)}(\lambda^2)&\equiv \int\limits_{k_{l}} \frac{1}{(k_{l}^2-\lambda^2)^{2}}
\ln^{l-1}{\left(-\frac{k_{l}^2-\lambda^2}{\lambda^2}\right)},
\end{align}
whose derivatives with respect to $\lambda$ can be straightforwardly obtained
\begin{align}
\lambda^2\frac{\partial I_{log}^{(l)}(\lambda^2)}{\partial \lambda^{2}}&=-(l-1)\, I_{log}^{(l-1)}(\lambda^2)- b \,\, (l-1)!\, \quad \mbox{where} \quad l> 1.
\end{align}
In the expressions above, at n-loop level,  $l$ can assume values from 1 to n. For simplicity, we also define $I_{log}^{(l=1)}(\lambda^2)=\ilog$.

This brief description is aimed to provide the reader with the most basic and necessary ingredients of the IREG rules, easing the discussion of the results of upcoming sections. For a more extensive review we refer to\cite{Arias-Perdomo:2021inz}.

Before presenting our examples though, we must tackle two important issues in the context of IREG: gauge invariance and the treatment of chiral theories. In some sense, they are intimately connected to each other as we discussed in \cite{Bruque:2018bmy}. To better illustrate this point, consider the 1-loop vacuum polarization tensor in massless QED whose amplitude is given by
\begin{equation}
    i\Pi_{\mu\nu}(p)=(-)(-ie)^{2}\int_{k}\Tr \Bigg\{ \gamma_{\mu} \dfrac{i}{(\slashed{k})} \gamma_{\nu} \dfrac{i}{(\slashed{k}-\slashed{p})} \Bigg\}.
\end{equation}

In order to comply with gauge invariance, one must always perform Dirac algebra before applying the IREG procedure we outlined before. Therefore, one obtains
\begin{equation}
    i\Pi_{\mu\nu}(p)=(-4e^{2})\int_{k}  \frac{2k^{\mu}k^{\nu}-k^{\mu}p^{\nu}-p^{\mu}k^{\nu}+g^{\mu\nu}(k.p-k^{2})}{k^{2}(k-p)^{2}}.
\end{equation}

Notice the appearance of a term with $k^2$ in the numerator. At first sight, one could be tempted to write $k^{2}=g^{\alpha\beta}k_{\alpha}k_{\beta}$. However, by requiring shift invariance, the following results follow 
\begin{equation}
\int_{k}  \dfrac{k^{2}}{k^{2}(k-p)^{2}}=\int_{k}\dfrac{1}{(k-p)^{2}}=\lim_{\mu^{2}\rightarrow 0}\int_{k}  \dfrac{1}{(k-p)^{2}-\mu^{2}}=\lim_{\mu^{2}\rightarrow 0} \int_{k}  \dfrac{1}{k^{2}-\mu^{2}}=0
\label{Nk2}
\end{equation}
\begin{small}
\begin{equation}
\begin{aligned}
g^{\alpha\beta}\int_{k}  \dfrac{k_{\alpha}k_{\beta}}{k^{2}(k-p)^{2}}
&=g^{\alpha\beta}\left\{\left(\frac{p_{\alpha}p_{\beta}}{3}-\frac{g_{\alpha\beta}p^{2}}{12}\right)\left[I_{\text{log}}(\lambda^{2})- b \ln\left(-\frac{p^2}{\lambda^2}\right) + \frac{13b}{6}\right]-\frac{g_{\alpha\beta}b p^{2}}{24}\right\}\\
&=-\frac{b p^{2}}{6}
\label{gab_kakb}
\end{aligned}
\end{equation}
\end{small}
which clearly shows that, in IREG, symmetric integration must be avoided in general 
	\beq
	\Bigg[\int_k k^{\mu_1}\cdots k^{\mu_{2m}} f(k^2)\Bigg]^{\text{ IREG}}\neq\;\frac{g^{\{\mu_1 \mu_2 } \cdots g^{\mu_{2m-1} \mu_{2m} \}}}{(2m)!} \Bigg[\int_k k^{2m} f(k^2)\Bigg]^{\text{ IREG}},
	\label{eq:sym}
	\eeq
where  the curly brackets indicate symmetrisation over Lorentz indices. In \cite{Bruque:2018bmy} a more extensive discussion can be found, leading to the following general statement: \textit{renormalization does not commute with index contraction}. Therefore, it is necessary to have some rules to consistently define a tensorial integral in IREG. We define normal form as the resulting integral after such rules were enforced. As exemplified in the case of massless QED, we  argue that, given a general Feynman amplitude, a normal form complying with gauge invariance is attained only after performing Dirac algebra and Lorentz contraction. In this way, terms such as $g_{\mu\nu}k^{2}$ or $k_{\mu}k_{\nu}$ are identified consistently and, since symmetric integration is not allowed, there is no risk of transforming one type of term into another. A pedagogical discussion of this point can also be found in \cite{Arias-Perdomo:2021inz}. Regarding chiral theories, a similar reasoning leads one to conclude that inadvertently using identities such as $\{\gamma_{\mu},\gamma_{5}\}=0$ may omit or change terms proportional to $k^{2}$. This results in a non-consistent procedure. A simple example is the evaluation of the following integral
\begin{align}\label{eq:first}
\int_{k}  \dfrac{\slashed{k}\gamma_{5}\slashed{k}}{k^{2}(k-p)^{2}}&
\overset{?}{=}
\frac{bp^{2}}{6}\gamma_{5},\quad\quad \mbox{using eq.\ref{gab_kakb} and $\{\gamma_{\mu},\gamma_{5}\}=0$,} \\
&\overset{?}{=}0, \quad\quad\quad\quad \mbox{using $\{\gamma_{\mu},\gamma_{5}\}=0$ and eq.\ref{Nk2}.}\label{eq:second}
\end{align}
At this point, one can ask which (if any) of the options above should be adopted in the context of IREG. To properly answer, we must adopt a more formal point of view. We start by mimicking 
the setup of dimensional methods where a set of Minkowski spaces is used.  For instance, in the case of DReg the method is defined in $QdS$,
$$
QdS= GnS \oplus Q(-2\epsilon)S,
$$
where $GnS$ is the genuine, four-dimensional space\footnote{We follow the same notation of \cite {Bruque:2018bmy}. Since in this contribution we only consider quantum field theories in four-dimensions, we could also use $G4S$.}. In this way, $\gamma_{5}$ belongs to $GnS$, while internal momenta, Dirac matrices, Minkowski metric, are defined in $QdS$. Similarly, DRED is defined in $QnS$
$$
QnS = QdS\oplus Q(2\epsilon)S= GnS \oplus Q(-2\epsilon)S\oplus Q(2\epsilon)S,
$$
with Dirac and Lorentz algebra defined in $QnS$ while the internal momenta are still defined in $QdS$.

In the case of IREG we adopt the same space $QnS$ of DRED. However, the setup is simpler since we can use $QnS$ as the space in which internal momenta live. For definiteness, in the case of IREG we have
$$
QnS= GnS \oplus X,
$$
where we define $X$ as an extra space which may not be realized explicitly.

Once the spaces are defined, we comment on passing some of their properties. We will define objects with $\bar{}\;$ as belonging to $GnS$, while a $\hat{}\;$ is reserved for space $X$. For instance, a momentum $p$ defined in $QnS$ can be split as $
p = \bar{p}+\hat{p}$. Thus, the identities below follow
\begin{align}
\{\gamma_{\mu},\gamma_{\nu}\}&=2g_{\mu\nu}\mathbb{1}; \quad \{\bar{\gamma}_{\mu},\bar{\gamma}_{\nu}\}=\{\gamma_{\mu},\bar{\gamma}_{\nu}\}=2\bar{g}_{\mu\nu}\mathbb{1}; \quad \gamma_{\mu}\gamma^{\mu}=\gamma_{\mu}\bar{\gamma}^{\mu}=4\;\mathbb{1}\\
\{\bar{\gamma}_{\mu},\hat{\gamma}_{\nu}\}&=0; \quad \{\gamma_{\mu},\hat{\gamma}_{\nu}\}=\{\hat{\gamma}_{\mu},\hat{\gamma}_{\nu}\}=2\hat{g}_{\mu\nu}\mathbb{1};\quad \gamma_{\mu}\hat{\gamma}^{\mu}=\bar{\gamma}_{\mu}\hat{\gamma}^{\mu}=\hat{\gamma}_{\mu}\hat{\gamma}^{\mu}=0.
\end{align}

Regarding $\gamma_{5}$ itself, we will adopt the definition
\beq
\gamma_{5}=-\frac{i}{4!}\epsilon_{abcd}\bar{\gamma}^{a}\bar{\gamma}^{b}\bar{\gamma}^{c}\bar{\gamma}^{d}
\label{eq:def G5}
\eeq
where we use Dirac matrices in $GnS$, consistently with $\gamma_{5}$ belonging to $GnS$. Given the definition above, the following properties are verified
\begin{align}
\{\bar{\gamma}_{\mu},\gamma_{5}\}=0; \quad \{\gamma_{\mu},\gamma_{5}\}=2\gamma_{5}\hat{\gamma_{\mu}}; \quad [\hat{\gamma}_{\mu},\gamma_{5}]=0
\end{align}
We call this the consistent scheme. For contrast, a naive scheme will enforce the anticommutativity of $\gamma_{5}$ with Dirac matrices defined in $QnS$.

Using the framework just exposed we can return to the previous integral
\begin{align}
\int_{k}  \dfrac{\slashed{k}\gamma_{5}\slashed{k}}{k^{2}(k-p)^{2}}&=
2\gamma_{5}\int_{k}  \dfrac{\hat{\slashed{k}}\slashed{k}}{k^{2}(k-p)^{2}}-\int_{k}  \dfrac{\gamma_{5}k^2}{k^{2}(k-p)^{2}}\nonumber\\
&=2\gamma_{5}\int_{k}  \dfrac{k^2}{k^{2}(k-p)^{2}}-2\gamma_{5}\int_{k}  \dfrac{\bar{k}^2}{k^{2}(k-p)^{2}}\nonumber\\
&=-2\bar{g}_{ab}\gamma_{5}\int_{k}  \dfrac{\bar{k}^{a}\bar{k}^{b}}{k^{2}(k-p)^{2}}=\frac{bp^{2}}{3}\gamma_{5}. \label{eq:third}
\end{align}
 
As can be seen, none of the previous proposals leads to the consistent one. The fact that it is actually the only consistent choice is exemplified in appendix \ref{ap:analytical}. Another important point to be noticed is that Lorentz contraction using the metric defined in $GnS$ does commute with renormalization, as already emphasized in \cite{Bruque:2018bmy}. Once the stage is set, we proceed in the following section to an application of the method and the ideas just exposed.

\section{Abelian Left-Model as an example of chiral theory}
\label{sec:QED} 


In this section we aim to exemplify how IREG can be consistently applied to a chiral theory. This will surpass the analysis of  \cite{Bruque:2018bmy} where we have applied the ideas of the last section in one-loop examples only. We have shown that, although consistent, IREG breaks BRST symmetry implying that finite BRST restoring-counterterms must be added (as it is well-known in the case of DReg). At higher orders, the inclusion of these terms is crucial to define a consistent renormalization program. This statement relies on the fact that, at one-loop, differences among the naive treatment of the $\gamma_{5}$ matrix and the consistent procedure can only amount to finite contributions after the limit to four-dimension is enforced (we provide an explicit analysis in the framework of IREG in appendix \ref{ap:dif}). Thus, as a step forward, we consider one of the simpler two-loop examples which may be affected by one-loop BRST restoring-counterterms: the gauge coupling $\beta$ function at two-loop level.


As working arena, we consider an abelian chiral model, denoted Abelian Left-Model, in which there is a set of massless left-handed fermionic field coupled to an abelian vector field, as below

\beq
\mathcal{L} = -\frac{1}{4}F_{\mu\nu}F^{\mu\nu}+\sum_{k}\bar{\psi_{k L}}\left[i\slashed{\partial} + eQ_{k} \slashed{A}\right]\psi_{k L} 
- \frac{1}{2\xi}(\partial^{\mu}A_{\mu})^{2}-\bar{c}\;\square \;c
\eeq
The charges of the fermionic fields are chosen in a way to render an anomaly free theory. Explicitly, they must fulfill $\sum_{k}Q_{k}^{3}=0$.  

This Lagrangian is initially defined in the physical (genuine) dimension, which means, in the notation introduced before, it is defined in $GnS$. It is easy to see that it is invariant under the BRST transformations
\begin{align}
\psi_{kL}(x) \rightarrow e^{iQ_{k}\theta c(x)}\psi_{kL}(x)\;, \quad   A_{\mu}(x) \rightarrow A_{\mu}(x) + \frac{\theta}{e}\partial_{\mu}c(x);\quad  
\bar{c}(x) \;\rightarrow \bar{c}(x) - \frac{\theta}{e}\frac{1}{\xi}\partial_{\mu}A^{\mu} 
\end{align}

We would like to define the regularized theory which implies that the Lagrangian (by consequence the Feynman rules) is defined in a different space. For DReg one must use $QdS$, while for IREG one adopts $QnS$. 
At first sight, this would be straightforward to be done, however, at closer inspection, some problems arise since we defined $\gamma_{5}$ in $GnS$. Given the hierarchy between the spaces, the fermionic propagator will mix terms in $GnS$ and $QnS$ being given by
\beq
\Delta(p) = \text{P}_{L}\frac{\slashed{p}}{\bar{p}^{2}}\text{P}_{R}
\eeq
where $\text{P}_{L/R}=(1\mp\gamma_{5})/2$, and $p$ lives in $QnS$ while $\bar{p}$ lives in $GnS$. This problem also happens if the theory is defined in $QdS$ \cite{Jegerlehner:2000dz,Belusca-Maito:2020ala}. In order to avoid this issue, we define the following Lagrangian in $QnS$
\beq
\mathcal{L}_{\text{QnS}} = -\frac{1}{4}F_{\mu\nu}F^{\mu\nu}+\sum_{k}\left[i\bar{\psi_{k}}\slashed{\partial}\psi_{k} + e Q_{k} \bar{\psi}_{kL}\slashed{A}\psi_{kL}\right] - \frac{1}{2\xi}(\partial^{\mu}A_{\mu})^{2}-\bar{c}\;\square \;c
\label{eq:lag1}
\eeq
which implies that a new field $\psi_{R}$ was implicitly introduced. The Lagrangian is no longer invariant under gauge transformations, the offending terms being given by
\beq
i\left(\bar{\psi}_{kL}\hat{\slashed{\partial}}\psi_{kR} +\bar{\psi}_{kR}\hat{\slashed{\partial}}\psi_{kL}\right) = i\bar{\psi_{k}}\left (\hat{\slashed{\partial}}\right)\psi_{k}
\eeq
where $\hat{\slashed{\partial}}$ lives in the $X$ space. From a practical point of view, the above terms will induce spurious anomalies which must be removed by the inclusion of finite counterterms. 
Finally, we would like to comment that the Lagrangian in $QnS$ can be recast in a axial-vector form where only the $\psi$ field appears
\begin{align}
\mathcal{L}_{\text{QnS}} &= -\frac{1}{4}F_{\mu\nu}F^{\mu\nu}+\sum_{k}\left[i\bar{\psi_{k}}\slashed{\partial}\psi_{k} + \frac{eQ_{k}}{2}\bar{\psi_{k}}\left[\bar{\gamma_{\mu}}\left(1-\gamma_{5}\right)\right]A^{\mu}\psi_{k}\right] - \frac{1}{2\xi}(\partial^{\mu}A_{\mu})^{2}-\bar{c}\;\square \;c\nonumber\\
&= -\frac{1}{4}F_{\mu\nu}F^{\mu\nu}+\sum_{k}\bar{\psi_{k}}\left[i\slashed{\partial} + eQ_{k}\left(\bar{\gamma_{\mu}}P_{L}\right)A^{\mu}\right]\psi_{k}  - \frac{1}{2\xi}(\partial^{\mu}A_{\mu})^{2}-\bar{c}\;\square \;c
\label{eq:lag}
\end{align}

Before moving to our examples, we emphasize that the analysis presented so far is only related to defining the Lagrangian in a different space than the genuine, four-dimensional space. As shown, this step is already non-trivial. A second step is the actual calculation of quantum corrections. As is well-known, the theory under consideration is anomaly free theory since we imposed that the charges fulfill $\sum_{k}Q_{k}^{3}=0$, so any anomaly that eventually appears is spurious and should be removed by finite symmetry-restoring counterterms. 

%
%

\subsection{1-loop diagrams}

We start our discussion at 1-loop level,
considering first the photon self-energy. The Feynman rules can be straightforwardly derived from Eq. \ref{eq:lag}, allowing one to obtain 
\begin{align}
i\Pi_{\mu\nu}=&- \frac{\sum_{j}Q_{j}^{2}e^{2}}{4} \int_{k} \text{Tr}\left[\frac{1}{\slashed{k}}\left(\bar{\gamma}_{\mu}-\bar{\gamma}_{\mu}\gamma_{5}\right)\frac{1}{\slashed{k}-\slashed{p}}\left(\bar{\gamma}_{\nu}-\bar{\gamma}_{\nu}\gamma_{5}\right)\right]\\
=&i\Pi_{\mu\nu}^{VV}+i\Pi_{\mu\nu}^{AV}+i\Pi_{\mu\nu}^{VA}+i\Pi_{\mu\nu}^{AA}
\end{align}


\noindent
Notice that the different types of fermionic fields do not mix, so their only effect is the sum over charges that appears as a global factor. In the equation above, the terms $i\Pi_{\mu\nu}^{AV}$ and $i\Pi_{\mu\nu}^{VA}$ contain an odd number of $\gamma_{5}$, so they will automatically vanish. It follows from the fact that, after using our definition of $\gamma_{5}$ and performing the Dirac trace, we obtain $\Pi_{\mu\nu}^{AV}\propto\epsilon_{\mu\nu \alpha \beta}p^{\alpha}p^{\beta}=0$ since the diagram contains two free Lorentz indexes and one external momenta only. Therefore, 
 we are left only with $\Pi_{\mu\nu}^{VV}$ and $\Pi_{\mu\nu}^{AA}$.

The first term will be given by
\begin{align}
i\Pi^{VV}_{\mu\nu} &=- \frac{\sum_{j}Q_{j}^{2}e^{2}}{4} \int_{k} \text{Tr}\left[\frac{1}{\slashed{k}}\bar{\gamma}_{\mu}\frac{1}{\slashed{k}-\slashed{p}}\bar{\gamma}_{\nu}\right]\nonumber\\
 &= -\sum_{j}Q_{j}^{2}e^{2} \int_{k} \frac{(k^{2}-k.p)\bar{g}_{\mu\nu}-2\bar{k}_{\mu}\bar{k}_{\nu}+\bar{p}_{\mu}\bar{k}_{\nu}+\bar{k}_{\mu}\bar{p}_{\nu}}{k^{2}(k-p)^{2}}
\end{align}
Using the IREG formalism, the end result is
\begin{align}
\frac{i\Pi^{VV}_{\mu\nu}}{\sum_{j}Q_{j}^{2}e^{2}}\Big|_{\text{IREG}}=&+\frac{1}{3}\left[\ilog-b\Log{p}+\frac{5}{3}b\right](\bb{g}_{\mu\nu}\bar{p}^{2}-\bb{p}_{\mu}\bb{p}_{\nu})\nonumber\\&-\frac{1}{3}\bb{g}_{\mu\nu}\hat{p}^{2}\left[\ilog-b\Log{p}+\frac{5}{3}b\right]
\end{align}
Notice the first line is just the result of standard QED (apart from global factors) if the metric and momenta were defined in $GnS$, while the second line contains a momentum defined in the $X$-space. The other term is more interesting 
\beq
i\Pi^{AA}_{\mu\nu} = - \frac{\sum_{j}Q_{j}^{2}e^{2}}{4} \int_{k} \text{Tr}\left[\frac{1}{\slashed{k}}\gamma_{5}\bar{\gamma}_{\mu}\frac{1}{\slashed{k}-\slashed{p}}\gamma_{5}\bar{\gamma}_{\nu}\right] = i\Pi^{VV}_{\mu\nu} + 2 \sum_{j}Q_{j}^{2}e^{2}\int_{k} \frac{(\hat{k}^{2}-\hat{k}.\hat{p})\bar{g}_{\mu\nu}}{k^{2}(k-p)^{2}}
\eeq
which amounts to
\begin{align}
\frac{i\Pi^{AA}_{\mu\nu}}{\sum_{j}Q_{j}^{2}e^{2}}\Big|_{\text{IREG}}=&\frac{i\Pi^{VV}_{\mu\nu}}{\sum_{j}Q_{j}^{2}e^{2}}\Big|_{\text{IREG}}+\frac{b}{3}\bar{g}_{\mu\nu}\bar{p}^{2}+\bar{g}_{\mu\nu}\hat{p}^{2}\left[\ilog-b\Log{p}+2b\right]\nonumber\\
\frac{i\Pi^{AA}_{\mu\nu}}{\sum_{j}Q_{j}^{2}e^{2}}\Big|_{\text{IREG}}=&+\frac{1}{3}\left[\ilog-b\Log{p}+\frac{5}{3}b\right](\bb{g}_{\mu\nu}\bar{p}^{2}-\bb{p}_{\mu}\bb{p}_{\nu})\nonumber\\&+\frac{b}{3}\bar{g}_{\mu\nu}\bar{p}^{2}+\frac{2}{3}\bb{g}_{\mu\nu}\hat{p}^{2}\left[\ilog-b\Log{p}+\frac{13}{6}b\right]
\end{align}
In this case, there is the appearance not only of a $X$-part, but there is a finite term which explictly breaks the gauge symmetry. The final result is:
\begin{align}
\frac{i\Pi_{\mu\nu}}{\sum_{j}Q_{j}^{2}e^{2}}\Big|_{\text{IREG}}=&+\frac{2}{3}\left[\ilog-b\Log{p}+\frac{5}{3}b\right](\bb{g}_{\mu\nu}\bar{p}^{2}-\bb{p}_{\mu}\bb{p}_{\nu})\nonumber\\&+\frac{b}{3}\bar{g}_{\mu\nu}\bar{p}^{2}+\frac{1}{3}\bb{g}_{\mu\nu}\hat{p}^{2}\left[\ilog-b\Log{p}+\frac{8}{3}b\right]
\label{eq:pimunu}
\end{align}
which is compatible with eq. 5.12 of \cite{Belusca-Maito:2020ala} and eq. C.1 of \cite{Martin:1999cc}.

The fermion self-energy for the fermionic field $\psi_{j}$ can be dealt in a similar way. Adopting Feynman gauge, one obtains 
\begin{align}
i\Sigma_{j}(-p)&= -\frac{Q_{j}^{2}e^{2}}{4} \int_{k} \left[\left(\bar{\gamma}_{\mu}-\bar{\gamma}_{\mu}\gamma_{5}\right)\frac{1}{\slashed{k}}\left(\bar{\gamma}^{\mu}-\bar{\gamma}^{\mu}\gamma_{5}\right)\right]\frac{1}{(k-p)^{2}}\\
i\Sigma_{j}(-p)\Big|_{\text{IREG}}&=-\frac{Q_{j}^{2}e^{2}}{2}\left[\ilog-b\Log{p}+2b\right](\slashed{\bar{p}}-\slashed{\bar{p}}\gamma_{5})\nonumber\\
&=-Q_{j}^{2}e^{2}\left[\ilog-b\Log{p}+2b\right]\slashed{\bar{p}}P_{L}
\label{eq:fermion}
\end{align}
In this case, apart from defining the gamma matrices at $GnS$, the result is the same of standard QED if DRED is applied \cite{Gnendiger:2017pys,Arias-Perdomo:2021inz}. 

Finally we consider the vertex $\bar{\psi}_{j}\gamma^{\mu}P_{L}A_{\mu}\psi_{j}$ whose amplitude (without external fields) we denote by $\Gamma_{j}^{\mu}$. Our main aim will be to discuss the Ward identity, so we will actually compute


\begin{align}
i(p_{1}+p_{2})_{\mu}\Gamma_{j}^{\mu}(p_{1},p_{2})&=Q_{j}^{3}e^{3}\int_{k}\bar{\gamma}_{\alpha}P_{L}\frac{1}{\slashed{k}-\slashed{p_{2}}}\left(\slashed{\bar{k}}+\slashed{\bar{p_{1}}}-\slashed{\bar{k}}+\slashed{\bar{p_{2}}}\right)P_{L}\frac{1}{\slashed{k}+\slashed{p_{1}}}\bar{\gamma}_{\alpha}P_{L}\frac{1}{k^{2}}\nonumber\\
&=iQ_{j}e\left[\Sigma(p_{1})-\Sigma(-p_{2})\right]+Q_{j}^{3}e^{3}\int_{k}\bar{\gamma}_{\alpha}P_{L}\frac{1}{\slashed{k}+\slashed{p_{1}}}\bar{\gamma}_{\alpha}P_{L}\frac{1}{k^{2}}\frac{(\hat{k}-\hat{p_{2}})^{2}}{(k-p_{2})^{2}}\nonumber\\
&-Q_{j}^{3}e^{3}\int_{k}\bar{\gamma}_{\alpha}P_{L}\frac{1}{\slashed{k}-\slashed{p_{2}}}\bar{\gamma}_{\alpha}P_{L}\frac{1}{k^{2}}\frac{(\hat{k}+\hat{p_{1}})^{2}}{(k+p_{1})^{2}}\nonumber\\
&=iQ_{j}e\left[\Sigma_{j}(p_{1})-\Sigma_{j}(-p_{2})\right]-Q_{j}^{3}e^{3}b(p_{1}+p_{2})^{\mu}\bar{\gamma}_{\mu}P_{L}
\label{ward}
\end{align}
As can be seen, the Ward identity is violated by a local finite term which is compatible with previous analysis \cite {Bruque:2018bmy,Martin:1999cc}. It should be noticed that, in the form just presented, it is clear that the breaking of the Ward identity comes from the mismatch between the fermion propagator (defined in $QnS$) and the vertex containing Dirac matrices in $GnS$, as exemplified in eq. \ref{eq:lag}. We should emphasize that we are considering an anomaly free theory by construction, so the breaking of the Ward identity is spurious and should be restored by the addition of finite symmetry-restoring counterterms. 


Before moving to two-loop order, we must discuss the renormalization of the theory we are considering. As can be seen from the 1-loop results we already presented, we need (at least) UV counterterms related to the following terms\footnote{Hereafter, an implicit summation in the index $k$ is to be understood.}
\beq 
 -\frac{1}{4}F_{\mu\nu}F^{\mu\nu}+i\bar{\psi}_{kL}\bar{\slashed{\partial}}\psi_{kL} + Q_{k}e \bar{\psi}_{kL}\slashed{A}\psi_{kL}.
\eeq
which could be achieved by performing multiplicative renormalization 
\beq
A_{\mu}\rightarrow Z_{A}^{1/2}A_{\mu},\quad \psi_{kL}\rightarrow Z_{L}^{1/2}\psi_{kL},\quad e\rightarrow Z_{e}e
\eeq
amounting to 
\begin{align}
\lag_{QnS}=-\frac{1}{4}Z_{A}F_{\mu\nu}F^{\mu\nu}+&iZ_{L}\bar{\psi}_{kL}\bar{\slashed{\partial}}\psi_{kL} + Q_{k}e Z_{L}Z_{A}^{1/2}Z_{e} \bar{\psi}_{kL}\slashed{A}\psi_{kL} + i\bar{\psi}_{kR}\bar{\slashed{\partial}}\psi_{kR}+
\nonumber\\
+&iZ_{L}^{1/2}\left(\bar{\psi}_{kL}\hat{\slashed{\partial}}\psi_{kR} +\bar{\psi}_{kR}\hat{\slashed{\partial}}\psi_{kL}\right)
\end{align}

Notice that we will get a UV counterterm for the purely X-part. Since the right-handed fermionic field does not couple to the gauge field, it does not pose any problem. This observation will prove to be crucial, since it will allow us to modify the terms containing the right-handed fermionic fields, without harmful consequences. On this regard, one should notice that the renormalized lagrangian just defined does not have the same functional form of eq. \ref{eq:lag1}. To recover this behavior, we have to consider some extra terms, proportional to $\psi_{kR}$, which must be added order-by-order. At one-loop order we can write $Z_{L}=1+\delta_{L}$, implying the renormalized Lagrangian can be recast as below
\begin{align}
\lag_{QnS}=&-\frac{1}{4}Z_{A}F_{\mu\nu}F^{\mu\nu}+i\bar{\psi}_{kL}\bar{\slashed{\partial}}\psi_{kL} + Q_{k}eZ_{A}^{1/2}Z_{e} \bar{\psi}_{kL}\slashed{A}\psi_{kL} + i\bar{\psi}_{kR}\bar{\slashed{\partial}}\psi_{kR}+
i\left(\bar{\psi}_{kL}\hat{\slashed{\partial}}\psi_{kR} +\bar{\psi}_{kR}\hat{\slashed{\partial}}\psi_{kL}\right)
\nonumber\\
&+i\delta_{L}\bar{\psi}_{kL}\bar{\slashed{\partial}}\psi_{kL} + Q_{k}e \delta_{e}Z_{A}^{1/2}Z_{e}\bar{\psi}_{kL}\slashed{A}\psi_{kL} 
+i\frac{\delta_{L}}{2}\left(\bar{\psi}_{kL}\hat{\slashed{\partial}}\psi_{kR} +\bar{\psi}_{kR}\hat{\slashed{\partial}}\psi_{kL}\right)
\label{eq:lagQ}
\end{align}

Notice that we have discarded terms at second order on $\delta_{L}$. In the format above, it is clear that the counterterm for the fermion propagator is not entirely defined in $QnS$. To recover this behavior, we can amend the renormalized Lagrangian with the following terms
\beq 
\lag_{amend}=i\frac{\delta_{L}}{2}\left(\bar{\psi}_{kL}\hat{\slashed{\partial}}\psi_{kR} +\bar{\psi}_{kR}\hat{\slashed{\partial}}\psi_{kL}\right)+i\delta_{L}\bar{\psi}_{kR}\bar{\slashed{\partial}}\psi_{kR}
\label{eq:amend}
\eeq
We emphasize, once again, that only terms proportional to $\psi_{kR}$ were added, which is allowed since the right-handed fermionic field decouples from the theory. 

On top of the amending UV counterterms, we must also add the finite symmetry-restoring terms
\beq 
\lag_{restore}=\delta_{f}^{V} Q_{k}e \bar{\psi}_{kL}\bar{\slashed{A}}\psi_{kL} + \delta_{f}^{A}A^{\mu}\bar {\partial}^{2}A_{\mu}
\eeq
where $\delta_{f}^{V} = Q_{k}^{2}e^{2}b$ and $\delta_{f}^{A} = 
\sum_{k}Q_{k}^{2}e^{2}b/3 $. The former is needed to restore the Ward identity given by eq. (\ref{ward}) while the latter is required to enforce transversality of $\Pi_{\mu\nu}$ when the limit to four dimensions is taken.

Finally, our renormalized  Lagrangian will be 
\begin{align}
\mathcal{L}_{\text{QnS}} = &-Z_{A}\frac{1}{4}F_{\mu\nu}F^{\mu\nu}+(1+\delta_{L})\left(i\bar{\psi}_{k}\slashed{\partial}\psi_{k} + Z_{e}Z_{A}^{1/2}Q_{k}e \bar{\psi}_{kL}\slashed{A}\psi_{kL} \right) \nonumber\\
&+ 
\delta_{f}^{V} Q_{k}e \bar{\psi}_{kL}\bar{\slashed{A}}\psi_{kL} + \delta_{f}^{A}A^{\mu}\bar {\partial}^{2}A_{\mu}
\label{eq:laq ren}
\end{align}

Focusing only in the terms with the UV counterterms, we have the same structure of eq. \ref{eq:lag}. Thus, as before, we have an evanescent gauge-breaking term, proportional to $ i\bar{\psi}\left (\hat{\slashed{\partial}}\right)\psi$. By construction, when the limit to four dimensions is taken we must recover gauge symmetry which implies the usual identity $Z_{e}Z_{A}^{1/2}=1$. Therefore, as in standard QED, we can obtain the gauge coupling beta function with knowledge only of the photon two-point function.


%

\subsection{2 loop diagrams}

%

\begin{figure}[h!]
\centering
\hfill
\subcaptionbox{}{\includegraphics[width=0.25\textwidth]{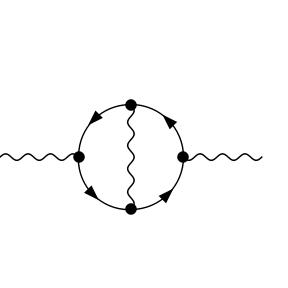}}%
\hfill
\subcaptionbox{}{\includegraphics[width=0.25\textwidth]{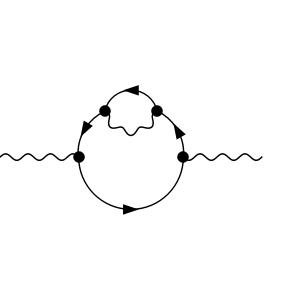}}%
\vfill
\caption{Two-loop correction to the two-point function of the photon field $A$.}
\label{QED:Feynman}
\end{figure}

In this section we present our main result: the calculation of the two-loop gauge coupling beta function for the Left-Model. As discussed in the end of the last section, the relation $Z_{e}=Z_{A}^{-1/2}$ holds, where $Z_{e}$, $Z_{A}$ are the renormalization constant of the gauge coupling and photon respectively. This simplifies the calculation since only the two-point function of the photon field will be required.  In particular, the diagrams to be considered are given in fig. \ref{QED:Feynman}.

We proceed to the details of the calculation.
Regarding the treatment of Dirac matrices, by adopting a naive $\gamma_{5}$ scheme, the terms involving none (vector), two (mixed) or four (axial) $\gamma_{5} $ will be identical apart from combinatorial global factors (as in the 1-loop case, only terms with an even number of $\gamma_{5}$ will survive). In table \ref{tab:naive} we organize the results obtained in the context of IREG, where $\mathcal{N}$ amounts to the combinatorial global factors we just mentioned. In the last line we provide the sum of all contributions. The coefficients A and B are defined as below
\begin{equation}
\label{eq: def A B}
\mathcal{A_{\mu\nu}}=\frac{i \sum_{j}Q_{j}^{4}e^{4}}{(4\pi)^{4}} \left[A \bar{g}_{\mu\nu}\bar{p}^{2} - B \bar{p}_{\mu}\bar{p}_{\nu}\right],
\end{equation}
where $\mathcal{A_{\mu\nu}}$ is the amplitude of the diagram under consideration.

\begin{table}[h!]
\begin{equation}
\begin{array}{|c|c|c|c|c|c|c|c|c|}
\hline
\multicolumn{1}{ |c }{\multirow{2}{*}{\text{Diagram}}}  & \multicolumn{4}{|c}{\text{A}} & \multicolumn{4}{|c|}{\text{B}} 
\\ \cline{2-9}
& I_{log}^{(2)}(\lambda^{2}) & I_{log}^{2}(\lambda^{2}) & \rho_{IREG} & I_{log}(\lambda^{2}) & I_{log}^{(2)}(\lambda^{2}) & I_{log}^{2}(\lambda^{2}) & \rho_{IREG} & I_{log}(\lambda^{2}) \\ \hline \hline
a \text{ (vector)}\quad\mathcal{N}=\frac{1}{16} & \frac{8}{3b} & -\frac{8}{3b^2} & \frac{8}{3b} & -\frac{140}{9b} & \frac{8}{3b} & -\frac{8}{3b^2} & \frac{8}{3b} & -\frac{128}{9b}
\\ \hline 
b \text{ (vector)}\quad\mathcal{N}=\frac{1}{16} & -\frac{8}{3b} & \frac{8}{3b^2} & -\frac{8}{3b} & \frac{104}{9b} &  -\frac{8}{3b} & \frac{8}{3b^2} & -\frac{8}{3b} & \frac{92}{9b}
\\ \hline
\text{Sum (vector)} & 0 & 0 & 0 & -\frac{1}{4b} & 0 & 0 & 0 & -\frac{1}{4b}
\\ \hline \hline
a \text{ (mixed)}\quad\mathcal{N}=\frac{6}{16} & \frac{8}{3b} & -\frac{8}{3b^2} & \frac{8}{3b} & -\frac{140}{9b} & \frac{8}{3b} & -\frac{8}{3b^2} & \frac{8}{3b} & -\frac{128}{9b}
\\ \hline 
b \text{ (mixed)}\quad\mathcal{N}=\frac{6}{16} & -\frac{8}{3b} & \frac{8}{3b^2} & -\frac{8}{3b} & \frac{104}{9b} &  -\frac{8}{3b} & \frac{8}{3b^2} & -\frac{8}{3b} & \frac{92}{9b}
\\ \hline 
\text{Sum (mixed)} & 0 & 0 & 0 & -\frac{3}{2b} & 0 & 0 & 0 & -\frac{3}{2b}
\\ \hline \hline
a \text{ (axial)}\quad\mathcal{N}=\frac{1}{16} & \frac{8}{3b} & -\frac{8}{3b^2} & \frac{8}{3b} & -\frac{140}{9b} & \frac{8}{3b} & -\frac{8}{3b^2} & \frac{8}{3b} & -\frac{128}{9b}
\\ \hline 
b \text{ (axial)}\quad\mathcal{N}=\frac{1}{16} & -\frac{8}{3b} & \frac{8}{3b^2} & -\frac{8}{3b} & \frac{104}{9b} &  -\frac{8}{3b} & \frac{8}{3b^2} & -\frac{8}{3b} & \frac{92}{9b}
\\ \hline 
\text{Sum (axial)} & 0 & 0 & 0 & -\frac{1}{4b} & 0 & 0 & 0 & -\frac{1}{4b}
\\ \hline \hline
\text{Sum (all)} & 0 & 0 & 0 & -\frac{2}{b} & 0 & 0 & 0 & -\frac{2}{b}
\\ \hline
\end{array}\nonumber
\end{equation}
\caption{Results using the naive scheme where $\rho_{IREG} = I_{log}(\lambda^{2})\log\left[-\frac{p^2}{\lambda^2}\right]$}
\label{tab:naive}
\end{table}


\newpage

For comparison, we reproduce in tab \ref{tab:qed} the results in standard QED \cite{Cherchiglia:2020iug}

\begin{table}[h!]
\begin{equation}
\begin{array}{|c|c|c|c|c|c|c|c|c|}
\hline
\multicolumn{1}{ |c }{\multirow{2}{*}{\text{Diagram}}}  & \multicolumn{4}{|c}{\text{A}} & \multicolumn{4}{|c|}{\text{B}} 
\\ \cline{2-9}
& I_{log}^{(2)}(\lambda^{2}) & I_{log}^{2}(\lambda^{2}) & \rho_{IREG} & I_{log}(\lambda^{2}) & I_{log}^{(2)}(\lambda^{2}) & I_{log}^{2}(\lambda^{2}) & \rho_{IREG} & I_{log}(\lambda^{2}) \\ \hline \hline
a & \frac{8}{3b} & -\frac{8}{3b^2} & \frac{8}{3b} & -\frac{140}{9b} & \frac{8}{3b} & -\frac{8}{3b^2} & \frac{8}{3b} & -\frac{128}{9b}
\\ \hline 
b & -\frac{8}{3b} & \frac{8}{3b^2} & -\frac{8}{3b} & \frac{104}{9b} &  -\frac{8}{3b} & \frac{8}{3b^2} & -\frac{8}{3b} & \frac{92}{9b}
\\ \hline 
\text{Sum} & 0 & 0 & 0 & -\frac{4}{b} & 0 & 0 & 0 & -\frac{4}{b}
\\ \hline  
\end{array}\nonumber
\end{equation}
\caption{Results for QED using IREG where $\rho_{IREG} = I_{log}(\lambda^{2})\log\left[-\frac{p^2}{\lambda^2}\right]$}
\label{tab:qed}
\end{table}

By comparing tables \ref{tab:naive} and \ref{tab:qed} one notices that the total divergent part differs by a factor $1/2$ which can be easily understood. In the Left-model we are considering, the Lagrangian can be written as in eq. \ref{eq:lag} where the coupling constant has a factor $1/2$. Therefore, the result in the naive scheme amounts to 
\begin{align}
\Pi_{\mu\nu}^{\text{naive}}=8\left(1/2\right)^{4}\Pi_{\mu\nu}^{\text{QED}}=\left(1/2\right)\Pi_{\mu\nu}^{\text{QED}}\;,
\end{align}
where the factor 8 comes from combinatorics while 1/2 comes from the Lagrangian (for simplicity, we are considering $\sum_{j}Q_{j}^{4}=1$).

On the other hand, by adopting our consistent procedure the results will be different, see table \ref{tab:bhmv}.
\begin{table}[h!]
\begin{equation}
\begin{array}{|c|c|c|c|c|c|c|c|c|}
\hline
\multicolumn{1}{ |c }{\multirow{2}{*}{\text{Diagram}}}  & \multicolumn{4}{|c}{\text{A}} & \multicolumn{4}{|c|}{\text{B}} 
\\ \cline{2-9}
& I_{log}^{(2)}(\lambda^{2}) & I_{log}^{2}(\lambda^{2}) & \rho_{IREG} & I_{log}(\lambda^{2}) & I_{log}^{(2)}(\lambda^{2}) & I_{log}^{2}(\lambda^{2}) & \rho_{IREG} & I_{log}(\lambda^{2}) \\ \hline \hline
a \text{ (vector)}\quad\mathcal{N}=\frac{1}{16} & \frac{8}{3b} & -\frac{8}{3b^2} & \frac{8}{3b} & -\frac{140}{9b} & \frac{8}{3b} & -\frac{8}{3b^2} & \frac{8}{3b} & -\frac{128}{9b}
\\ \hline 
b \text{ (vector)}\quad\mathcal{N}=\frac{1}{16} & -\frac{8}{3b} & \frac{8}{3b^2} & -\frac{8}{3b} & \frac{104}{9b} &  -\frac{8}{3b} & \frac{8}{3b^2} & -\frac{8}{3b} & \frac{92}{9b}
\\ \hline
\text{Sum (vector)} & 0 & 0 & 0 & -\frac{1}{4b} & 0 & 0 & 0 & -\frac{1}{4b}
\\ \hline \hline
a \text{ (mixed)}\quad\mathcal{N}=\frac{6}{16} & \frac{8}{3b} & -\frac{8}{3b^2} & \frac{8}{3b} & -\frac{164}{9b} & \frac{8}{3b} & -\frac{8}{3b^2} & \frac{8}{3b} & -\frac{152}{9b}
\\ \hline 
b \text{ (mixed)}\quad\mathcal{N}=\frac{6}{16} & -\frac{8}{3b} & \frac{8}{3b^2} & -\frac{8}{3b} & \frac{104}{9b} &  -\frac{8}{3b} & \frac{8}{3b^2} & -\frac{8}{3b} & \frac{92}{9b}
\\ \hline 
\text{Sum (mixed)} & 0 & 0 & 0 & -\frac{7}{12b} & 0 & 0 & 0 & -\frac{7}{12b}
\\ \hline\hline
a \text{ (axial)}\quad\mathcal{N}=\frac{1}{16} & \frac{8}{3b} & -\frac{8}{3b^2} & \frac{8}{3b} & -\frac{188}{9b} & \frac{8}{3b} & -\frac{8}{3b^2} & \frac{8}{3b} & -\frac{176}{9b}
\\ \hline 
b \text{ (axial)}\quad\mathcal{N}=\frac{1}{16} & -\frac{8}{3b} & \frac{8}{3b^2} & -\frac{8}{3b} & \frac{104}{9b} &  -\frac{8}{3b} & \frac{8}{3b^2} & -\frac{8}{3b} & \frac{92}{9b}
\\ \hline 
\text{Sum (axial)} & 0 & 0 & 0 & -\frac{5}{2b} & 0 & 0 & 0 & -\frac{5}{2b}
\\ \hline\hline
\text{Sum (all)} & 0 & 0 & 0 & -\frac{10}{3b} & 0 & 0 & 0 & -\frac{10}{3b}
\\ \hline
\end{array}\nonumber
\end{equation}
\caption{Results for the consistent procedure where $\rho_{IREG} = I_{log}(\lambda^{2})\log\left[-\frac{p^2}{\lambda^2}\right]$}
\label{tab:bhmv}
\end{table}

As can be readily noticed, terms proportional to $I_{log}^{(2)}(\lambda^{2})$, $I_{log}^{2}(\lambda^{2})$, $\rho_{IREG}$ are all identical to the naive $\gamma_{5}$ scheme. This behavior was expected since differences among the naive $\gamma_{5}$ scheme and our procedure amount to finite local terms at 1-loop level. Therefore, only local terms proportional to $I_{log}(\lambda^{2})$ can be affected, as ours results show. Moreover, the vector contributions are identical to standard QED (apart from global factors) as expected as well.
Regarding the sum of all contributions we notice: 
1) terms proportional to $I_{log}^{(2)}(\lambda^{2})$, $I_{log}^{2}(\lambda^{2})$ cancel as in standard QED (in the context of dimensional methods, it implies that no $\epsilon^{-2} $ term will survive); 2) coefficients A and B are the same, implying the UV divergent part of the two-loop two-point function of the photon is transverse; 3) the coefficient of $I_{log}(\lambda^{2})$ is \textbf{not} the same as in table \ref{tab:naive}. The second point
is an consequence of discarding terms with evanescent momentum $\hat{p}$ when writing the results using the consistent procedure. It is justified since $p$ is the physical momentum of the photon, which must be taken to four dimensions in the end of the calculation. A similar reasoning could be applied to the 1-loop photon two-point function given by eq. \ref{eq:pimunu} if one is performing  an 1-loop calculation.
Finally, the third point could imply a distinct value for the gauge coupling $\beta$ function. However, as we are going to see, when the finite symmetry-restoring counterterms are added \textbf{and} the 1-loop UV counterterms are included, the same result of table \ref{tab:naive} will be obtained. 

We begin with the analysis of the 1-loop counterterms. For the calculation we are interested at, only the 1-loop divergences of the vertex and fermion self-energy will be needed, which (apart from global factors and the left projector) are identical to standard QED. The situation would be different if the 1-loop photon self-energy was necessary as well, since in this case there is a new divergent term proportional to an evanescent contribution. Nevertheless, when the 1-loop counterterms are inserted they will amount to a non-vanishing result as opposed to standard QED. The reason boils down to the mismatch between the fermion propagator (defined in $QnS$) and the fermion self-energy 1-loop correction proportional to $\slashed{\bar{p}}$. This behavior will be corrected by adding the counterterms due to the  amending terms.
%
 After this brief discussion, we proceed to the details of the calculation. We begin with the counterterm related to the electron self-energy, that comes from eq. \ref{eq:fermion}
\begin{align}
i\Pi_{\mu\nu}^{\Sigma}&=\sum_{j}Q_{j}^{2}e^{2}\int_{k} \text{Tr}\left\{\frac{1}{\slashed{k}}\;\left[-\Sigma^{\text{div}}(k)\right]\;\frac{1}{\slashed{k}}\bar{\gamma}_{\mu}P_{L}\frac{1}{\slashed{k}-\slashed{p}}\bar{\gamma}_{\nu}P_{L}\right\} \nonumber\\
&= -\sum_{j}Q_{j}^{4}e^{4}\int_{k} \text{Tr}\left\{\frac{1}{\slashed{k}}\,\left[\ilog\slashed{\bar{k}}P_{L}\right]
 \,\frac{1}{\slashed{k}}\bar{\gamma}_{\mu}P_{L}\frac{1}{\slashed{k}-\slashed{p}}\bar{\gamma}_{\nu}P_{L}\right\}\nonumber\\
 &= -\sum_{j}Q_{j}^{4}e^{4}\int_{k} \text{Tr}\left\{
\left(1-\frac{\hat{k}^{2}}{k^{2}}\right)\,\left[\ilog \right]
 \,\frac{1}{\slashed{k}}\bar{\gamma}_{\mu}P_{L}\frac{1}{\slashed{k}-\slashed{p}}\bar{\gamma}_{\nu}P_{L}\right\}
\end{align}

As anticipated, the mismatch between the fermion propagator and $\Sigma$ generates a term with $\hat{k}^{2}$ which will prevent the cancellation of the counterterms, as can be seen by considering the counterterm containing the vertex
\begin{align}
i\Pi_{\mu\nu}^{\Gamma}&=-\sum_{j}Q_{j}^{2}e^{2}\int_{k} \text{Tr}\left\{\frac{1}{\slashed{k}}\;\left[-\Gamma_{\mu}^{\text{div}}\right]\;\frac{1}{\slashed{k}-\slashed{p}}\bar{\gamma}_{\nu}P_{L}\right\} \nonumber\\
&= -\sum_{j}Q_{j}^{4}e^{4}\int_{k} \text{Tr}\left\{\frac{1}{\slashed{k}}\,\left[-\ilog\bar{\gamma}_{\mu}P_{L}\right]\,\frac{1}{\slashed{k}-\slashed{p}}\bar{\gamma}_{\nu}P_{L}\right\}\nonumber\\
&= -\sum_{j}Q_{j}^{4}e^{4}\int_{k} \text{Tr}\left\{\left[-\ilog\right]\frac{1}{\slashed{k}}\,\bar{\gamma}_{\mu}P_{L}\,\frac{1}{\slashed{k}-\slashed{p}}\bar{\gamma}_{\nu}P_{L}\right\}
\end{align}
which is just the result from the 1-loop photon self-energy given by eq. \ref{eq:pimunu} multiplied by $\ilog$.

Therefore, the sum of the 1-loop counterterm (responsible for the subtraction of subdivergences) gives
\begin{align}
i\Pi_{\mu\nu}^{\text{CT}}=2i\Pi_{\mu\nu}^{\Sigma}+2i\Pi_{\mu\nu}^{\Gamma}&= -2\sum_{j}Q_{j}^{4}e^{4}\int_{k} \text{Tr}\left\{
\frac{\hat{k}^{2}}{k^{2}}\,\left[-\ilog \right]
 \,\frac{1}{\slashed{k}}\bar{\gamma}_{\mu}P_{L}\frac{1}{\slashed{k}-\slashed{p}}\bar{\gamma}_{\nu}P_{L}\right\}\nonumber\\
 &=\frac{i\sum_{j}Q_{j}^{4}e^{4}}{(4\pi)^{4}}\left[\frac{2}{3b}\ilog\left(\bb{g}_{\mu\nu}\bar{p}^{2}-\bb{p}_{\mu}\bb{p}_{\nu}\right)\right]
\end{align}
where the factor two comes from the multiplicity of the diagrams.  

Turning to the amending counterterms, coming from eq. \ref{eq:lagQ} and the last term of eq. \ref{eq:amend}, one obtains (once again, the global factor 2 comes from the multiplicity of the diagram)
\begin{align}
i\Pi_{\mu\nu}^{\text{amend}}&=2\sum_{j}Q_{j}^{2}e^{2}\int_{k} \text{Tr}\left\{\frac{1}{\slashed{k}}\;\left[-\Sigma^{\text{amend}}\right]\frac{1}{\slashed{k}}\bar{\gamma}_{\mu}P_{L}\;\frac{1}{\slashed{k}-\slashed{p}}\bar{\gamma}_{\nu}P_{L}\right\} \nonumber\\
&= -2\sum_{j}Q_{j}^{4}e^{4}\int_{k} \text{Tr}\left\{\frac{1}{\slashed{k}}\,\left[\ilog\;\slashed{\hat{k}}+\ilog\;\slashed{\bar{k}}P_{R}\right]
 \,\frac{1}{\slashed{k}}\bar{\gamma}_{\mu}P_{L}\frac{1}{\slashed{k}-\slashed{p}}\bar{\gamma}_{\nu}P_{L}\right\}\nonumber\\
 &= -2\sum_{j}Q_{j}^{4}e^{4}\int_{k} \text{Tr}\left\{
\left(\frac{\hat{k}^{2}}{k^{2}}\right)\;\left[\ilog\right]\;\frac{1}{\slashed{k}}\bar{\gamma}_{\mu}P_{L}\frac{1}{\slashed{k}-\slashed{p}}\bar{\gamma}_{\nu}P_{L}\right\}
\end{align} 
As can be seen, the contribution of $i\Pi_{\mu\nu}^{\text{amend}}$ exactly cancel those from $i\Pi_{\mu\nu}^{\text{CT}}$. This is expected since $i\Pi_{\mu\nu}^{\text{CT}}$ was non-null only due to a mismatch between the fermion propagator and the fermion counterterm, which the amending terms by construction corrected. 

Regarding the finite symmetry restoring counterterms they can be obtained by eq. \ref{ward} 
\begin{align}
i\Pi_{\mu\nu}^{\text{rest}}&=-2\sum_{j}Q_{j}e\int_{k} \text{Tr}\left\{\frac{1}{\slashed{k}}\;\Gamma_{\mu}^{\text{rest}}\;\frac{1}{\slashed{k}-\slashed{p}}\bar{\gamma}_{\nu}P_{L}\right\} \nonumber\\
&=-2\sum_{j}Q_{j}^{4}e^{4} \int_{k} \text{Tr}\left\{\frac{1}{\slashed{k}}\,\left[b\;\bar{\gamma}_{\mu}P_{L}\right]\,\frac{1}{\slashed{k}-\slashed{p}}\bar{\gamma}_{\nu}P_{L}\right\}\nonumber\\
&=\frac{i\sum_{j}Q_{j}^{4}e^{4}}{(4\pi)^{4}}\left[\frac{4}{3b}\ilog\left(\bb{g}_{\mu\nu}\bar{p}^{2}-\bb{p}_{\mu}\bb{p}_{\nu}\right)\right]
\end{align}
where the global factor 2 comes from the multiplicity of the diagram.

Therefore, the final contribution that must be added to $\Pi_{\mu\nu}$ is the sum of $\Pi_{\mu\nu}^{\text{CT}}$, $\Pi_{\mu\nu}^{\text{amend}}$ and $\Pi_{\mu\nu}^{\text{rest}}$ amounting to 
\begin{align}
i\Pi_{\mu\nu}^{\text{total}}=&\frac{i \sum_{j}Q_{j}^{4}e^{4}}{(4\pi)^{4}} \left( g_{\mu\nu}p^{2} - p_{\mu}p_{\nu}\right)\left(-\frac{10}{3b}+\frac{2}{3b}-
\frac{2}{3b}+\frac{4}{3b}\right)\ilog\nonumber\\
=&\frac{i \sum_{j}Q_{j}^{4}e^{4}}{(4\pi)^{4}} \left( g_{\mu\nu}p^{2} - p_{\mu}p_{\nu}\right)\left(-\frac{2}{b}
\right)\ilog
\end{align}

As can be seen, we recover the \textbf{same} result of the naive scheme, which implies that the two-loop $\beta$ function is \textbf{not} modified when our consistent procedure is enforced. Such result is reassuring of previous analysis performed in the literature based on the naive scheme, which is not consistent in general. 

To conclude, we present the $\beta$ function of the gauge coupling in our model. As standard, we define 
\begin{align}
\beta=\lambda^{2}\frac{d}{d \lambda^{2}} \frac{\alpha}{\pi}=-\frac{\alpha}{\pi}\lambda^{2}\frac{d}{d \lambda^{2}} \ln Z_{\alpha}=\frac{\alpha}{\pi}\lambda^{2}\frac{d}{d \lambda^{2}} \ln Z_{A}  
\end{align}
where $Z_{\alpha}=Z_{e}^{2}/4\pi$, and $Z_{A}$ is given by
\begin{align}
Z_{A}&=1+\frac{\alpha}{4\pi}\left(-\frac{\sum_{j}2Q_{j}^{2}}{3b}\ilog\right)+\frac{\alpha^{2}}{(4\pi)^{2}}\left(-\frac{\sum_{j}2Q_{j}^{4}}{b}\ilog\right)
\end{align}
Finally, using eq. \ref{eq:derivative}, we obtain 
\beq
\beta = \left[\frac{\sum_{j}8Q_{j}^{2}}{3} \left(\frac{\alpha}{4\pi}\right)^{2} + \sum_{j}8Q_{j}^{4}\left(\frac{\alpha}{4\pi}\right)^{3}\right]
\eeq

We should emphasize that the coefficients for the $\beta$ function given above were obtained applying two different procedures. First we used a naive $\gamma_{5}$ scheme in the context of IREG. Afterwards, we used the consistent version of IREG, in which the hierarchy between Minkowski spaces is enforced. In any of the cases, the same final result was attained.

\section{Conclusions}
\label{sec:conclusions} 

In the upcoming years, with the high-luminosity program of LHC moving forward, it will be fundamental to push the precision frontier further. From a theoretical point of view, it will be necessary to have better control both from non-perturbative and perturbative aspects. The former requires an increase knowledge of PDFs, with good prospects in the following years \cite{Cruz-Martinez:2021rgy}. The latter concerns the development of innovative techniques to handle multiloop calculations, which may encompass the use of different regularization methods. Particularly promising are techniques that stay, as much as possible, in four-dimension. Among these, the implicit regularization method, defined in momentum space, is tailored to extract the UV divergence of generic higher loop integrals in a way to comply with unitary, locality and Lorentz invariance. It improves in the BPHZ regularization by complying with abelian gauge invariance in general as well as non-abelian gauge in working examples. 

In this context, another important issue is to which extend the regularization method can be consistently applied in chiral theories. This problem is well-known to dimensional techniques, however, only recently it was shown that similar problems occur in regularization methods defined in the physical dimension that comply with gauge invariance. In this contribution we extent the analysis of \cite{Bruque:2018bmy} to two-loop order, using an abelian chiral model as working example. We obtained the two-loop gauge coupling beta function using a naive $\gamma_{5}$ scheme as well as the consistent procedure first envisaged in \cite{Bruque:2018bmy}. The results agree which reassure the usage of the naive scheme even though it is inconsistent in general. The subtleties involved in the calculation using the consistent procedure are discussed. As perspective, we aim to apply the consistent procedure to an non-abelian theory in general.

\section*{Acknowledgments}

I gratefully acknowledge enlightening discussions with Dominik Stockinger, M. Pérez-Victoria, Brigitte Hiller and Marcos Sampaio. The author acknowledges support from National Council for
Scientific and Technological Development – CNPq through the project 166523/2020-8 and Fundação para a Ciência e Tecnologia (FCT) through the project CERN/FIS-COM/0035/2019. This publication is based upon work from COST Action CA16201 PARTICLEFACE, supported by COST (European Cooperation in Science and Technology, www.cost.eu).

\appendix

\section{Exemplifying the consistent scheme}
\label{ap:analytical} 
%
%
%
%
%

In this appendix we review some results of \cite{Bruque:2018bmy} to exemplify the method. For simplicity we will consider a 2-dimensional euclidean space and compute two-point functions at one-loop. We will be mainly concerned with the VA (vector-axial) correlator, which allows to discuss vector and axial Ward identities in perfect analogy to the well-known AVV anomaly in four-dimensions. This correlator is given by

\begin{align}
\Pi^5_{\mu\nu}(p) & = -\int \frac{d^2 k}{4\pi^2} \, \text{tr} \left(\gamma_{\mu}\frac{1}{\slashed{k}-\slashed{p}}\gamma_{\nu}\gamma_{5}\frac{1}{\slashed{k}} \right) = - \text{tr}\left(\gamma_{\mu}\gamma_\alpha\gamma_{\nu}\gamma_{5} \gamma_{\beta}\right) B_{\alpha\beta}.
\end{align}

In section \ref{sec:ireg}, we discussed some procedures to deal with the $\gamma_{5}$ matrix. The first choice (eq. \ref{eq:first}) was to perform the regularization and only after perform Dirac algebra. Following this procedure, we just need the integral $B_{\alpha\beta}(p)$ evaluated within IREG
\begin{align}
B_{\alpha\beta}(p)  =  \int \frac{d^2 k}{4\pi^2} \, \frac{(k-p)_\alpha k_\beta}{k^2(k-p)^2} = \frac{1}{4\pi} \left\{\frac{\delta_{\alpha\beta}}{2} \left(\ilog-\log \frac{p^2}{\lambda^2}\right) + \left(\delta_{\alpha\beta}- \frac{p_\alpha p_\beta}{p^2}\right) \right\}. \label{Bmunu}
\end{align}
For the trace, we use the definition of $\gamma_{5}$ in 2 dimensions to obtain
\beq
\text{tr}\left(\gamma_{\mu}\gamma_\alpha\gamma_{\nu}\gamma_{5} \gamma_{\beta}\right)=
2\left(-\epsilon_{\beta\nu}\delta_{\alpha\mu}+
\epsilon_{\mu\nu}\delta_{\alpha\beta}-\epsilon_{\alpha\nu}\delta_{\beta\mu}+\epsilon_{\beta\alpha}\delta_{\mu\nu}
-\epsilon_{\mu\alpha}\delta_{\beta\nu}-\epsilon_{\beta\mu}\delta_{\alpha\nu}\right). \label{tracefour5}
\eeq
The end result is
\beq
\Pi^5_{\mu\nu}(p)  = - \frac{1}{2\pi}\left(\epsilon_{\mu\nu}-\frac{2\epsilon_{\alpha\nu}p_{\mu}p_{\alpha}}{p^{2}}\right)
\eeq
By computing the Ward identities, one readily obtain
\begin{align}
p_{\mu}\Pi^5_{\mu\nu}(p)&= \frac{1}{2\pi}\epsilon_{\mu\nu}p_{\mu},\nn
p_{\nu}\Pi^5_{\mu\nu}(p)&= \frac{1}{2\pi}\epsilon_{\nu\mu}p_{\nu},
\end{align}
showing that the anomaly is equally distributed among the vectorial and axial Ward identity. Since we want to have the vectorial Ward identity respected, our first choice to deal with the $\ga_5$ matrix is not adequate. 

The second choice (eq. \ref{eq:second}) was to use identities such as $\{\ga_\mu,\ga_5\}=0$ to simplify the integrand, and then apply IREG. It is enlightening to consider the Ward identities directly in this case. For the vectorial one,
\begin{align}
p_\mu \Pi^5_{\mu\nu}(p) &= - \int \frac{d^2 k}{4\pi^2} \, \text{tr} \left((\slashed{p}- \slashed{k}+\slashed{k})\frac{1}{\slashed{k}-\slashed{p}}\ga_{\nu}\gamma_{5}\frac{1}{\slashed{k}} \right)  \nn
&= \int \frac{d^2 k}{4\pi^2} \, \text{tr} \left(\ga_{\nu}\gamma_{5}\frac{1}{\slashed{k}}\right) -\int \frac{d^2 k}{4\pi^2} \, \text{tr} \left(\frac{1}{\slashed{k}-\slashed{p}}\ga_{\nu}\gamma_{5} \right) \nn
&=0\label{eq:vector}
\end{align}
where we have performed the shift $k\rightarrow k+p$ in the second integral of the second line. For the axial Ward identity
\begin{align}
p_\nu \Pi^5_{\mu\nu}(p) &= - \int \frac{d^2 k}{4\pi^2} \, \text{tr} \left(\gamma_{\mu}\frac{1}{\slashed{k}-\slashed{p}}(\slashed{p}- \slashed{k}+\slashed{k})\gamma_{5}\frac{1}{\slashed{k}} \right) \nn
&=  \int \frac{d^2 k}{4\pi^2} \, \text{tr} \left(\ga_{\mu}\gamma_{5}\frac{1}{\slashed{k}}\right) + \int \frac{d^2 k}{4\pi^2} \, \text{tr} \left(\gamma_{\mu}\frac{1}{\slashed{k}-\slashed{p}}\gamma_{5} \right)\nn
&=0
\end{align}
where we have used $\{\ga_\mu,\ga_5\}$ in the second integral integral of the second line, and used the same identity again to obtain the third line. As can be seen, the second proposal respects the vectorial Ward identity, but also cannot reproduce the anomaly. 

Therefore, we are left with the third proposal, which  is the consistent scheme. In this case, the Dirac matrices other than $\ga_5$ are defined in $QnS$, and we have the identity  $\{\gamma_{\mu},\gamma_{5}\}=2\gamma_{5}\hat{\gamma_{\mu}}$. Thus, while the vectorial Ward identity can be evaluated just as above (eq.\ref{eq:vector}), the axial Ward identity will give
\begin{align}
p_\nu \Pi^5_{\mu\nu}(p) &= - \int \frac{d^2 k}{4\pi^2} \, \text{tr} \left(\gamma_{\mu}\frac{1}{\slashed{k}-\slashed{p}}(\slashed{p}- \slashed{k}+\slashed{k})\gamma_{5}\frac{1}{\slashed{k}} \right)  \nn
&= \int \frac{d^2 k}{4\pi^2} \, \text{tr} \left(\ga_{\mu}\gamma_{5}\frac{1}{\slashed{k}}\right) + \int \frac{d^2 k}{4\pi^2} \, \text{tr} \left(\gamma_{\mu}\frac{1}{\slashed{k}-\slashed{p}}\gamma_{5} \right) - 2 \int \frac{d^2 k}{4\pi^2} \, \text{tr} \left(\gamma_{\mu}\frac{1}{\slashed{k}-\slashed{p}}
 \gamma_5 \hat{\slashed{k}}\frac{1}{\slashed{k}} \right) \nn
&= \int \frac{d^2 k}{4\pi^2} \, \text{tr} \left(\ga_{\mu}\gamma_{5}\frac{1}{\slashed{k}}\right) + \int \frac{d^2 k}{4\pi^2} \, \text{tr} \left(\gamma_{\mu}\frac{1}{\slashed{k}-\slashed{p}}\gamma_{5} \right) - 2 \int \frac{d^2 k}{4\pi^2} \, \text{tr} \left(\gamma_{\mu}\frac{1}{\slashed{k}-\slashed{p}}
 \gamma_5 \frac{k^{2}-\bar{k^{2}}}{k^{2}} \right)\nn
&= \int \frac{d^2 k}{4\pi^2} \, \text{tr} \left(\ga_{\mu}\gamma_{5}\frac{1}{\slashed{k}}\right) - \int \frac{d^2 k}{4\pi^2} \, \text{tr} \left(\gamma_{\mu}\frac{1}{\slashed{k}-\slashed{p}}\gamma_{5} \right) + 2 \int \frac{d^2 k}{4\pi^2} \, \text{tr} \left(\gamma_{\mu}\frac{1}{\slashed{k}-\slashed{p}}
 \gamma_5 \frac{\bar{k^{2}}}{k^{2}} \right)\nn
&=0+\frac{\epsilon_{\nu\mu}p_{\nu}}{\pi}
\end{align}
where we have used a similar reasoning to eq. \ref{eq:third}, performed the shift $k\rightarrow k + p$ in some of the integrals as well as used the fact that odd integral in k are null. As can be seen, using the consistent scheme, the vectorial Ward identity is respected and the anomaly is correctly reproduced. 

\section{Differences among the naive and consistent procedure at one-loop}
\label{ap:dif}

In this appendix we discuss what differences can one expect between the naive and consistent procedure of $\gamma_{5}$ at one-loop order in the framework of IREG. Consider, for instance, the one-loop amplitude below
\begin{equation}
F \equiv\int_{k} A_{\mu_{1}\cdots\mu_{n}}(k)B_{\nu_{1}\cdots\nu_{m}}(k) = F_{\text{div}}+F_{\text{fin}},  
\end{equation} 
where $A$, $B$ may depend on external momenta as well, and contain Dirac matrices. For simplicity we omit the Lorentz indexes in the function $F$. We are assuming IREG regularization in order to isolate the UV divergent part from the finite contribution. If we insert a $\gamma_{5}$ between the functions $A$, $B$, we obtain within the naive treatment
\begin{equation}
F^{5}_{N} \equiv\int_{k} A_{\mu_{1}\cdots\mu_{n}}(k)\gamma_{5}B_{\nu_{1}\cdots\nu_{m}}(k) = (F_{\text{div}}+F_{\text{fin}})(-1)^{l}\gamma_{5},  
\end{equation} 
where $l$ is the number of Dirac matrices contained in the function $B$. 

In the consistent procedure, the calculation is more involved. For simplicity, we consider the case that the function $B$ contains only one Dirac matrix. In this situation, we can split the $B$ function into two parts, each belonging to $GnS$ or the $X$-space, allowing us to rewrite $F^{5}$ as below
\begin{align}
F^{5}_{C} &\equiv\int_{k} A_{\mu_{1}\cdots\mu_{n}}(k)\gamma_{5}\left[\bar{B}_{\nu_{1}\cdots\nu_{m}}(k)+\hat{B}_{\nu_{1}\cdots\nu_{m}}(k)\right] \nonumber\\
&= \int_{k} A_{\mu_{1}\cdots\mu_{n}}(k)\left[-\bar{B}_{\nu_{1}\cdots\nu_{m}}(k)+\hat{B}_{\nu_{1}\cdots\nu_{m}}(k)\right]\gamma_{5}\nonumber\\
&= \int_{k} A_{\mu_{1}\cdots\mu_{n}}(k)\left[-B_{\nu_{1}\cdots\nu_{m}}(k)+2\hat{B}_{\nu_{1}\cdots\nu_{m}}(k)\right]\gamma_{5}\nonumber\\
&= (F_{\text{div}}+F_{\text{fin}})(-1)\gamma_{5} + 2\int_{k} A_{\mu_{1}\cdots\mu_{n}}(k)\hat{B}_{\nu_{1}\cdots\nu_{m}}(k)\gamma_{5}.  
\end{align}
Thus, the difference between the two prescriptions for the $\gamma_{5}$ matrix is given by
\begin{align}
F^{5}_{C} - F^{5}_{N} &= 2\int_{k} A_{\mu_{1}\cdots\mu_{n}}(k)\hat{B}_{\nu_{1}\cdots\nu_{m}}(k)\gamma_{5},\nonumber\\
&=2\int_{k} A_{\mu_{1}\cdots\mu_{n}}(k)\left[B_{\nu_{1}\cdots\nu_{m}}(k)-\bar{B}_{\nu_{1}\cdots\nu_{m}}(k)\right]\gamma_{5}.  
\end{align} 
In the case that only one external momentum is present, we obtain the result 
\begin{equation}
F^{5}_{C} - F^{5}_{N} = (p_{\alpha}-\bar{p}_{\alpha})G + (g_{\beta\alpha}-\bar{g}_{\beta\alpha})H + \text{similar terms},
\end{equation}
where $G$, $H$ are functions obtained after regularization, $\beta$ is a Lorentz index belonging to the set $\{\mu_{1},\cdots,\mu_{n},\nu_{1},\cdots,\mu_{m}\}$, and $\alpha$ is the Lorentz index related to the Dirac matrix contained in the function $B$. As can be seen, if the limit to four-dimensions is now enforced, we explicitly obtain a null result. This conclusion is expected within the framework of IREG, since there is only one Lorentz index belonging to $GnS$ in this case. Explicitly, there is no ambiguity in obtaining the result  
\begin{equation}
\int_{k} A_{\mu_{1}\cdots\mu_{n}}(k)B_{\nu_{1}\cdots\nu_{m}}(k)\Big|_{\text{IREG}} = p_{\alpha}G + g_{\beta\alpha}H + \text{similar terms}. 
\end{equation}
If one of the Lorentz indexes (for instance, $\alpha$) is now defined in $GnS$, one can just modify the end result, replacing $p_{\alpha}\rightarrow\bar{p}_{\alpha}$, and $g_{\beta\alpha}\rightarrow \bar{g}_{\beta\alpha}$.
A similar analysis can be performed for the case with more than one Dirac matrix contained in the function $B$, and/or more external momenta present. The important point here is that all Dirac matrices contain free Lorentz indexes, as opposed to the situation exemplified with eq. \ref{eq:third}. 

Finally, if contractions are present that lead to terms proportional to $k^{2}$, care must be exercised. In this case, the difference among the naive treatment and the consistent procedure will be related to 
\begin{align}
\int_{k} \left[k^{2} - \bar{k}^{2}\right]f(k) = 
\int_{k} \left[k^{2} - \bar{g}_{\alpha\beta}k^{\alpha}k^{\beta}\right]f(k), 
\end{align}  
where $f$ may depend on Lorentz indexes and/or external momenta. Using the framework of IREG, for the UV divergent part, any external momenta will only appear in the numerator. Thus, we obtain
\begin{align}
 \int_{k} \left[k^{2} - \bar{k}^{2}\right]f(k)\Big|_{\text{div}}= \int_{k} \left[k^{2} - \bar{g}_{\alpha\beta}\frac{g^{\alpha\beta}k^{2}}{4}\right]g(k)= 0   
\end{align}
where $g(k)$ depends only on $k^{2}$. The above result shows that there is no difference for the UV divergent part among the two procedures in this case. However, the finite contribution will be different in general, as already exemplified within eq. \ref{eq:third}.

\end{document}